\RequirePackage{silence}
\documentclass[twocolumn,fleqn,usenatbib,useAMS]{aastex631} 
\acceptjournal{Astronomy}
\usepackage{graphicx}
\usepackage{amsmath}

\DeclareMathOperator{\acos}{acos}
\DeclareMathOperator{\acosh}{acosh}
\usepackage{amsfonts}
\usepackage{float}
\usepackage{bm}
\usepackage{ae,aecompl}
\usepackage{array}
\usepackage{soul}
\usepackage{mathtools}
\usepackage{multirow}
\usepackage[utf8]{inputenc}
\usepackage{booktabs}
\usepackage{graphicx}

\DeclareRobustCommand{\appropto}{\mathrel{\vcenter{
    \offinterlineskip\halign{\hfil$##$\cr 
        \propto\cr\noalign{\kern2pt}\sim\cr\noalign{\kern-2pt}}}}}

\shorttitle{Constraints on $H_0$ and $\Omega_{\mathrm{M}}$ without modelling the CMB anisotropies}
\shortauthors{I. Banik and N. Samaras}

\begin{document}

\title{Constraints on the Hubble and matter density parameters with and without modelling the CMB anisotropies}


\author[0000-0002-4123-7325]{Indranil Banik}
\affiliation{Institute of Cosmology \& Gravitation, University of Portsmouth, Dennis Sciama Building, Burnaby Road, Portsmouth PO1 3FX, UK}

\author[0000-0001-8375-6652]{Nick Samaras}
\affiliation{Astronomical Institute, Faculty of Mathematics and Physics, Charles University, V Hole\v{s}ovi\v{c}k\'ach 2, CZ-180 00 Praha 8, Czech Republic}

\correspondingauthor{Indranil Banik}
\email{indranil.banik@port.ac.uk}

\begin{abstract}
We consider constraints on the Hubble parameter $H_0$ and the matter density parameter $\Omega_{\mathrm{M}}$ from: (i) the age of the Universe based on old stars and stellar populations in the Galactic disc and halo \citep{Cimatti_2023}; (ii) the turnover scale in the matter power spectrum, which tells us the cosmological horizon at the epoch of matter-radiation equality \citep{Philcox_2022}; and (iii) the shape of the expansion history from supernovae (SNe) and baryon acoustic oscillations (BAOs) with no absolute calibration of either, a technique known as uncalibrated cosmic standards \citep*[UCS;][]{Lin_2021_UCS}. A narrow region is consistent with all three constraints just outside their $1\sigma$ uncertainties. Although this region is defined by techniques unrelated to the physics of recombination and the sound horizon then, the standard \emph{Planck} fit to the CMB anisotropies falls precisely in this region. This concordance argues against early-time explanations for the anomalously high local estimate of $H_0$ (the `Hubble tension'), which can only be reconciled with the age constraint at an implausibly low $\Omega_{\mathrm{M}}$. We suggest instead that outflow from the local KBC supervoid \citep*{Keenan_2013} inflates redshifts in the nearby universe and thus the apparent local $H_0$. Given the difficulties with solutions in the early universe, we argue that the most promising alternative to a local void is a modification to the expansion history at late times, perhaps due to a changing dark energy density.

\end{abstract}


\section{Introduction}

The Universe is thought to have undergone an early period of rapid accelerated expansion known as inflation, which would cause any initial departure from flatness to rapidly decay \citep{Guth_1981}. After this inflationary era and the subsequent era of radiation domination, a flat FLRW universe \citep{Friedmann_1922, Friedmann_1924, Lemaitre_1931, Robertson_1935, Robertson_1936a, Robertson_1936b, Walker_1937} is largely governed by just two parameters: the fraction of the cosmic critical density in the matter component ($\Omega_{\mathrm{M}}$) and $h$, the Hubble constant $H_0$ in units of 100~km/s/Mpc. It is well known that if we assume the standard cosmological paradigm known as Lambda-Cold Dark Matter \citep*[$\Lambda$CDM;][]{Peebles_1984, Efstathiou_1990, Ostriker_Steinhardt_1995}, $h$ and $\Omega_{\mathrm{M}}$ can be obtained to high precision from the anisotropies in the cosmic microwave background \citep[CMB; e.g.][]{Planck_2020}.\footnote{The CMB also provides a wealth of other information, including the fraction $\Omega_{\mathrm{b}}$ of the cosmic critical density in baryons, the optical depth to reionisation, and the power spectrum of primordial density fluctuations.} However, the inferred $h$ is inconsistent with the locally inferred value from the redshift gradient $z' \equiv dz/dr$, which measures how quickly the redshift $z$ rises with distance $r$. In a homogeneously expanding universe, the local $z' = H_0/c$, where $c$ is the speed of light \citep*[equation~3 of][]{Mazurenko_2025}. The anomalously high local $cz'$ compared to the predicted $H_0$ is known as the Hubble tension \citep[e.g.][]{Perivolaropoulos_2024, Valentino_2025}.

Although most local measurements of $cz'$ in recent years exceed the $\Lambda$CDM prediction calibrated using the CMB, some slightly lower measurements are plausibly in agreement with it, highlighting that the Hubble tension is still a matter of active investigation \citep{Efstathiou_2020, Freedman_2020, Boruah_2021, Freedman_2021, Lin_2021_inconsistency, Chen_2024_SNe, Freedman_2025}. Some studies constrain the expansion history using measurements at higher $z$. These $H_0$ measurements are usually in line with the $\Lambda$CDM prediction calibrated to the CMB \citep{Dominguez_2019, Birrer_2020, Park_2020, Cao_2022, Wu_2022}. This raises the possibility of obtaining different $H_0$ values from data at different $z$, as indeed suggested by several recent studies \citep{Jia_2023, Jia_2025a, Jia_2025b}. We explore this important issue further.

It has been argued that $h$ and $\Omega_{\mathrm{M}}$ are some of the most important cosmological parameters to consider when trying to address the Hubble tension \citep{Toda_2024}. Some proposed solutions modify the physics prior to recombination at $z = 1100$ but have minimal impact beyond the first Myr of cosmic history. This motivates us to infer $h$ and $\Omega_{\mathrm{M}}$ without modelling the CMB anisotropies in $\Lambda$CDM $-$ such results may not be applicable to alternative cosmological models, especially those involving early dark energy \citep[EDE; e.g.][]{Poulin_2019}. We therefore focus on three non-CMB constraints, two of which are almost completely immune to the physics in the first Myr. The third assumes $\Lambda$CDM prior to and shortly after matter-radiation equality at $z = 3400$. This is because early-time solutions to the Hubble tension often modify the physics only shortly prior to recombination, thus not much affecting the radiation-dominated era when the universe was $\gtrsim 7\times$ younger (see Appendix~\ref{a_matter_radiation}). All three constraints should therefore remain valid in such alternative cosmologies.

In Section~\ref{Methods}, we explain the techniques used to constrain $H_0$ and $\Omega_{\mathrm{M}}$. We then present our results and discuss them in Section~\ref{Results} before concluding in Section~\ref{Conclusions}.

\section{Methods}
\label{Methods}

In what follows, we briefly describe each of the three considered non-CMB constraints on $\Omega_{\mathrm{M}}$ and $h$, treating the local $cz'$ as a constraint on the latter. In each case, we consider two different studies that obtain the relevant constraint in different ways, increasing confidence in the robustness of our result. For completeness, we also consider the standard \emph{Planck} constraint, which assumes $\Lambda$CDM in the early universe \citep{Planck_2020}.

\subsection{Uncalibrated cosmic standards (UCS)}
\label{UCS}

There is much controversy about the absolute magnitude of Type~Ia supernovae (SNe~Ia) and the comoving size $r_d$ of the BAO ruler. Leaving these as free parameters in the UCS technique, we can still constrain the shape of the expansion history and thus $\Omega_{\mathrm{M}}$ \citep*{Lin_2021_UCS}.\footnote{Those authors also use the angular scale of the first peak in the CMB power spectrum as an extra angular BAO data point, but this does not assume any pre-recombination model. The comoving size of the CMB ruler is $r_\star = r_d/1.0184$, a ratio which is essentially the same even in quite non-standard cosmological models \citep[see section~2.2 of][]{Vagnozzi_2023}.} This yields the tight constraint $\Omega_{\mathrm{M}} = 0.302 \pm 0.008$.

It has recently become possible to use BAO measurements from the Dark Energy Spectroscopic Instrument Data Release 2 (DESI~DR2) to constrain $\Omega_{\mathrm{M}}$ without making assumptions about $r_d h$, which can be inferred from the data jointly with $\Omega_{\mathrm{M}}$ \citep{DESI_2025}. Their equation~17 shows that $\Omega_{\mathrm{M}} = 0.2975 \pm 0.0086$ without considering SNe or any other external dataset. BAO measurements are mostly sensitive around $z \approx 0.5-1$, thus probing epochs long after recombination.

\subsection{The matter power spectrum}
\label{LSS}

The cosmic expansion history has largely been a power-law due to matter being by far the dominant component, apart from the relatively recent phenomenon of dark energy dominance. The combination of a power-law primordial power spectrum \citep{Harrison_1970, Peebles_1970, Zeldovich_1972} with a power-law expansion history leads to a power-law spectrum also at late times. This behaviour breaks down on small scales because at the early times when these modes entered the cosmic horizon, the universe was still dominated by radiation. During that era, even sub-horizon matter density perturbations could not grow due to the Meszaros effect \citep{Meszaros_1974}. This leads to a characteristic peak in the matter power spectrum at a comoving wavenumber of $k_{\mathrm{eq}}$.

Surveys of large-scale structure (LSS) in redshift space provide the tight constraint $k_{\mathrm{eq}}/h = \left( 1.64 \pm 0.05 \right) \times 10^{-2}$/Mpc \citep{Philcox_2022}. This is important to our discussion because $k_{\mathrm{eq}}/h \propto \Omega_{\mathrm{M}} h$ \citep[equation~3 of][note they use $\Omega_0$ to denote $\Omega_{\mathrm{M}}$]{Eisenstein_1998}. We obtain the proportionality constant by noting that \citet{Philcox_2022} report most likely values of $h = 0.648$ and $\Omega_{\mathrm{M}} = 0.338$ (see their figure~2). This combination presumably corresponds to their most likely $k_{\mathrm{eq}}/h$. We assume its fractional uncertainty is the same as that in $\Omega_{\mathrm{M}} h$. We use their study only to constrain this product rather than follow their approach of breaking the degeneracy using a prior on $\Omega_{\mathrm{M}}$, since this often comes from the same information already considered in Section~\ref{UCS}, which we need to avoid double-counting.

The degeneracy between $\Omega_{\mathrm{M}}$ and $h$ can be broken by considering the matter power spectrum in greater detail instead of just focusing on where it peaks. By cross-correlating \emph{Planck} CMB observations from \emph{Planck} Public Release 4 \citep*[PR4;][]{Carron_2022} with galaxies observed as part of the Atacama Cosmology Telescope Data Release 6 \citep[ACT~DR6;][]{Mathew_2024, Qu_2024}, it is possible to build up three different two-point galaxy correlation functions, namely the galaxy-galaxy, galaxy-lensing, and lensing-lensing correlations.\footnote{Lensing here refers to weak lensing.} This $3 \times 2$pt analysis of LSS constrains both $h$ and $\Omega_{\mathrm{M}}$ reasonably precisely, albeit at the expense of requiring that structure growth on the relevant scales follow $\Lambda$CDM expectations \citep{Farren_2025}. Those authors combine these measurements with uncalibrated SNe~Ia to tighten the constraints.

\subsection{Stellar ages in the Galactic halo}
\label{Age_constraint}

\citet{Cimatti_2023} compile reliable measurements of the ages of stars, globular clusters (GCs), and ultrafaint dwarf galaxies in the Galactic disc and halo. The inverse variance weighted mean age of their 11 considered samples is
\begin{eqnarray}
    t_{\mathrm{age}} = 14.05 \pm 0.25 \, \text{Gyr},
\end{eqnarray}
with the inverse of the combined variance found by summing the individual inverse variance of each sample.

To estimate how long these objects took to form, we note that those authors estimated the formation redshift to be $11-30$, which corresponds to $t_{\mathrm{f}} = 100-400$~Myr after the Big Bang. Taking the geometric mean of this range, we estimate that $t_{\mathrm{f}} = 0.2$~Gyr, so it is uncertain by a factor of 2. The total age of the Universe is thus
\begin{eqnarray}
    t_{\mathrm{U}} ~=~ t_{\mathrm{age}} + t_{\mathrm{f}}.
    \label{t_U}
\end{eqnarray}

Since $t_{\mathrm{f}} \ll t_{\mathrm{age}}$, the total $t_{\mathrm{U}}$ is barely affected even by rather large changes to the assumed $t_{\mathrm{f}}$. For instance, if the expansion rate at $z > 30$ was actually 10\% faster than assumed such that the Universe then had an age of only 90~Myr rather than 100~Myr, the impact on $t_{\mathrm{U}}$ would be $<0.1\%$.

As a second way to estimate $t_{\mathrm{U}}$ independently of cosmological assumptions, we consider the work of \citet{Valcin_2025}, who focus exclusively on Galactic GCs. These should provide more reliable results than individual stars, but since the number of GCs is much smaller, it is more difficult to reliably determine the upper limit to their age distribution. \citet{Valcin_2025} fit this distribution using a truncated Gaussian and find that the upper limit to the GC ages is $13.39 \pm 0.25$~Gyr, combining statistical and systematic uncertainties in quadrature. The somewhat lower value may reflect that GCs took longer to form than stars, but we continue to make the assumption that $t_{\mathrm{f}} = 0.2$~Gyr and has a factor of 2 uncertainty. It will become apparent that assuming a higher $t_{\mathrm{f}}$ would only strengthen our main conclusion.

We obtain $H_0 t_{\mathrm{U}}$ for each $\Omega_{\mathrm{M}}$ by inverting equation~45 of \citet*{Haslbauer_2020}. We note that $H_0 t_{\mathrm{U}} \appropto \Omega_{\mathrm{M}}^{-0.28}$ \citep*[equation~35 of][]{Poulin_2023}.

\subsection{The \texorpdfstring{\emph{Planck}}{Planck} CMB constraint in \texorpdfstring{$\Lambda$}{L}CDM}

We find the size and orientation of the \emph{Planck} error ellipse from the published uncertainties on $\Omega_{\mathrm{M}}$ and $h$ \citep[see the TTTEEE column of table~5 in][]{Tristram_2024}. The tightest constrained combination is $\Omega_{\mathrm{M}} h^3$ \citep*{Kable_2019}. Working in the space of $\left( \ln \Omega_{\mathrm{M}}, \ln h \right)$, the short axis of the error ellipse is thus towards the direction $\left(1, 3 \right)/\sqrt{10}$, with uncertainty $\sigma_+$ (named for its positive slope). The long axis is the orthogonal direction $\left(-3, 1 \right)/\sqrt{10}$, with uncertainty $\sigma_-$. Decomposing the $\ln \Omega_{\mathrm{M}}$ and $\ln h$ directions into combinations of these two statistically independent directions, we find $\sigma_+$ and $\sigma_-$ by solving
\begin{eqnarray}
    \begin{bmatrix}
        0.1 & ~0.9 \\
        0.9 & ~0.1
    \end{bmatrix}
    \begin{bmatrix}
        \sigma_+^2 \\
        \sigma_-^2
    \end{bmatrix}
    ~=~
    \begin{bmatrix}
        \sigma^2 \left( \ln \Omega_{\mathrm{M}} \right) \\
        \sigma^2 \left( \ln h \right)
    \end{bmatrix},
\end{eqnarray}
where $\sigma \left( \ln \Omega_{\mathrm{M}} \right)$ is the fractional uncertainty in $\Omega_{\mathrm{M}}$ (similarly for $h$). We then scale the ellipse by a factor $\chi$ so that the enclosed probability is 68.27\%, corresponding to the usual $1\sigma$ definition. Since there are two degrees of freedom and we assume Gaussian errors, we find that $\chi = 1.52$ by inverting equation~24 of \citet*{Asencio_2021}. Thus, the error ellipse in $\ln$-space has semi-minor axis $\chi \sigma_+ = 0.0024$ and semi-major axis $\chi \sigma_- = 0.036$. The axis ratio of 15 indicates negligible uncertainty on $\Omega_{\mathrm{M}} h^3$ compared to other combinations.


These small uncertainties assume no major systematics with \emph{Planck} measurements of the $\mathcal{O} \left( 10^{-5} \right)$ CMB anisotropies. These seem quite robust given that the inferred cosmological parameters in $\Lambda$CDM remain similar if we use instead CMB measurements from ACT \citep{Calabrese_2025} or the South Pole Telescope \citep[SPT;][]{Camphuis_2025}. Moreover, even if we were to neglect measurements of the CMB anisotropies, we can infer $H_0$ using only the CMB monopole temperature $T_0$ in combination with the primordial light element abundances, relying on the well-established theory of Big Bang nucleosynthesis (BBN) to get the ratio $\eta$ between the number densities of baryons and photons \citep{Cyburt_2016}. Combining $\eta$ with the radiation density implied by $T_0$ tells us the baryon density $\Omega_{\mathrm{b}} h^2$. This in turn allows us to calculate $r_d$ given $\Omega_{\mathrm{M}}$ and $h$ \citep*{Brieden_2023}. Assuming $\Lambda$CDM physics sets the sound speed prior to recombination, their equation~3.4 shows that $r_d \appropto h^{-0.46}$ at fixed $\Omega_{\mathrm{b}} h^2$ and $\Omega_{\mathrm{M}}$. Bearing in mind that uncalibrated BAO data can already constrain $\Omega_{\mathrm{M}}$ and the product $r_d h$ (Section~\ref{UCS}), combining with BBN constrains $r_d$ and $h$ individually \citep{Schoneberg_2022}. This BAO + BBN route to $H_0$ gives $68.51 \pm 0.58$~km/s/Mpc \citep{DESI_2025}, far below local measurements of $cz'$. Therefore, any problem with measurements of the subtle CMB anisotropies cannot by itself account for the Hubble tension.

\subsection{The local redshift gradient}
\label{Local_redshift_gradient}

As a representative of the high local estimates of $cz'$ (usually equated with $H_0$), we show the SH0ES result \citep{Breuval_2024}. Their study calibrates the Leavitt Law linking the period and peak absolute magnitude of Cepheid variables \citep{Leavitt_1912} using geometric distances to Galactic Cepheids, both Magellanic Clouds, and NGC~4258. The Leavitt Law is then used to calibrate distances to host galaxies of SNe~Ia, whose absolute magnitude is important to determining $cz'$ \citep{Chen_2024_SNe}.

We emphasize that many other measurements of the local $cz'$ by different teams using different techniques and instruments give quite similar values \citep[e.g., see figure~10 of][and references therein]{Riess_2024}. Neither Cepheids nor SNe~Ia are critical to measuring the high local $cz'$, and indeed one can avoid using either \citep{Scolnic_2023}. By measuring the distance scale using Cepheid variables, the tip of the red giant branch, and surface brightness fluctuations, \citet{Uddin_2024} found that $H_0 \approx 72 - 73$~km/s/Mpc. Moreover, distances obtained by the \emph{James Webb Space Telescope} (\emph{JWST}) are in good agreement with those obtained previously by the \emph{Hubble Space Telescope} \citep[HST;][]{Riess_2024_consistency, Freedman_2025}, demonstrating that unrecognised crowding of Cepheid variables is not responsible for distances being underestimated as much as the \emph{Planck} cosmology requires \citep{Riess_2024_crowding}. The low $cz'$ reported by \citet{Freedman_2025} is due to a chance fluctuation caused by their small sample size and intrinsic scatter in the absolute magnitudes of SNe~Ia, but extending their results by including other galaxies with available data leads to a higher inferred $cz'$ in line with SH0ES \citep{Riess_2024_consistency, Li_2025}.

We consider a second estimate of the local $cz'$ using the fundamental plane \citep[FP;][]{Dressler_1987} relation of elliptical galaxies observed by DESI. The FP has recently been used to anchor the local $cz'$ to $d_{\mathrm{Coma}}$, the distance to the Coma Cluster \citep{Said_2025}. Measurements of $d_{\mathrm{Coma}}$ over the last three decades using a variety of techniques give consistent results, which combined with the DESI~FP result gives a high local $cz'$ consistent with other studies \citep{Scolnic_2025}. $cz'$ can be reduced to the \emph{Planck} $\Lambda$CDM value only if we assume an implausibly large $d_{\mathrm{Coma}} \gtrsim 110$~Mpc, but observations show that $d_{\mathrm{Coma}} \approx 100$~Mpc. This makes it increasingly unlikely that the local $cz'$ will one day be revised downwards by the necessary 10\% and local distance moduli revised upwards by 0.2~mag.

\section{Results and Discussion}
\label{Results}

\begin{figure*}
    \centering
    \includegraphics[width = \linewidth]{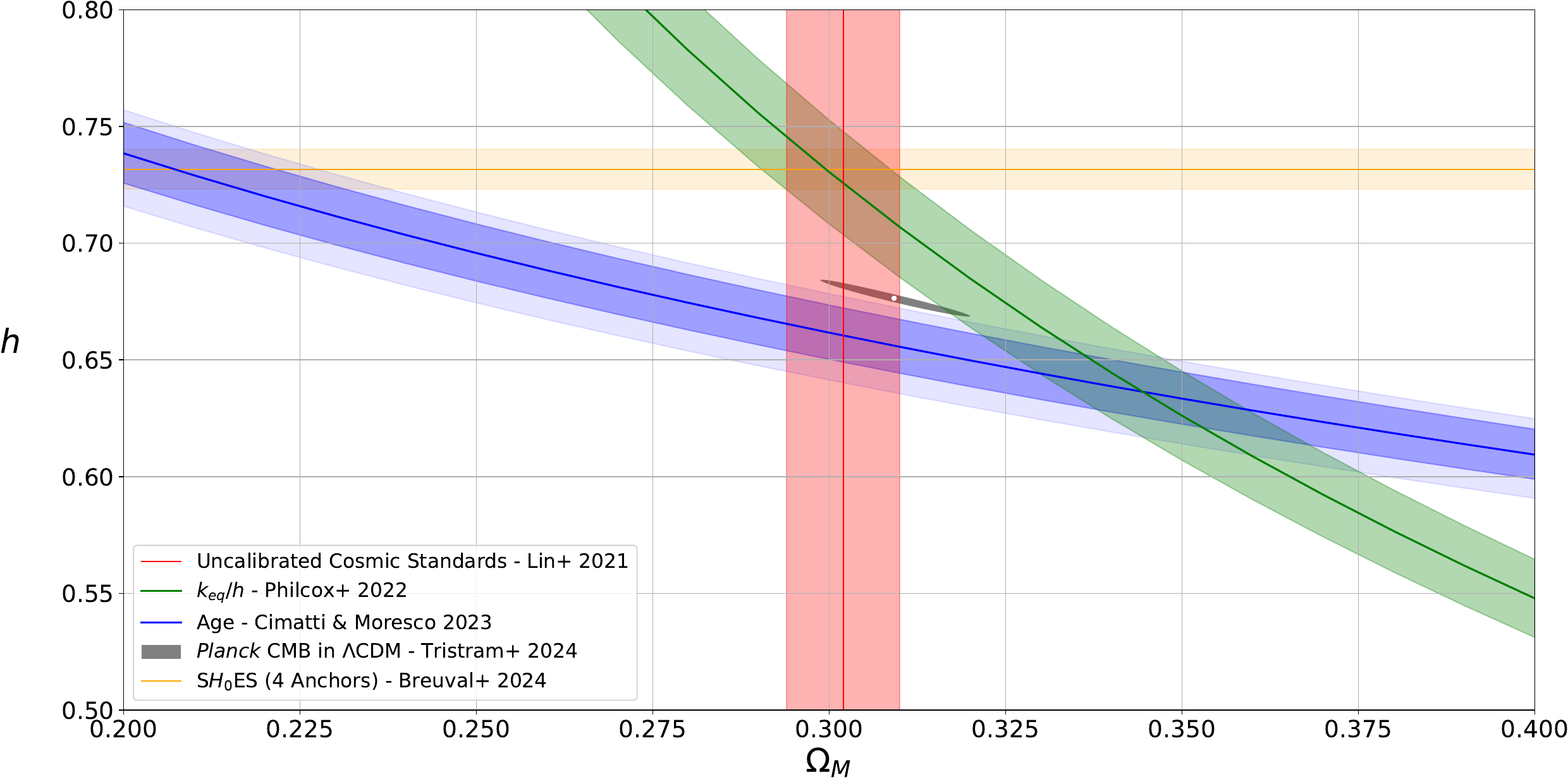}
    \caption{The $1\sigma$ constraints on $\Omega_{\mathrm{M}}$ and $h \equiv H_0$ in units of 100~km/s/Mpc from the shape of the expansion history traced by SNe and BAOs with no absolute calibration of either \citep*[red;][]{Lin_2021_UCS}, the turnover scale in the matter power spectrum \citep[green;][]{Philcox_2022}, and the ages of old stars in the Galactic disc and halo \citep[blue;][]{Cimatti_2023}, with the light blue band allowing a factor of 2 uncertainty in their formation time. The grey error ellipse shows the \emph{Planck} fit to the CMB anisotropies in $\Lambda$CDM \citep{Tristram_2024}, which provide the tightest constraint on the combination $\Omega_{\mathrm{M}} h^3$ \citep*{Kable_2019}. The white dot at its centre shows the most likely values. The yellow band shows $h$ estimated from the local redshift gradient by the SH0ES team \citep{Breuval_2024}, with 4 anchor galaxies used to calibrate the Leavitt Law.}
    \label{OmegaM_h}
\end{figure*}

Our results are shown in Figure~\ref{OmegaM_h}. The $1\sigma$ constraints from UCS, $k_{\mathrm{eq}}/h$, and stellar ages are not mutually compatible. However, these bands enclose a narrow triangle in parameter space that is consistent with all three constraints just outside their $1\sigma$ regions. Remarkably, this is precisely where we find the tight \emph{Planck} constraint from the CMB anisotropies assuming $\Lambda$CDM. The overlap of these two very small regions of parameter space strongly suggests that the underlying assumptions are largely correct and that the model parameters $\Omega_{\mathrm{M}}$ and $h$ are in this narrow range. The only outlier is the SH0ES result for the local $cz'$ using 42 SNe~Ia with distances found from the Leavitt Law, itself calibrated using 4 anchor galaxies with trigonometric distances (Section~\ref{Local_redshift_gradient}).

\begin{figure*}
    \centering
    \includegraphics[width = \linewidth]{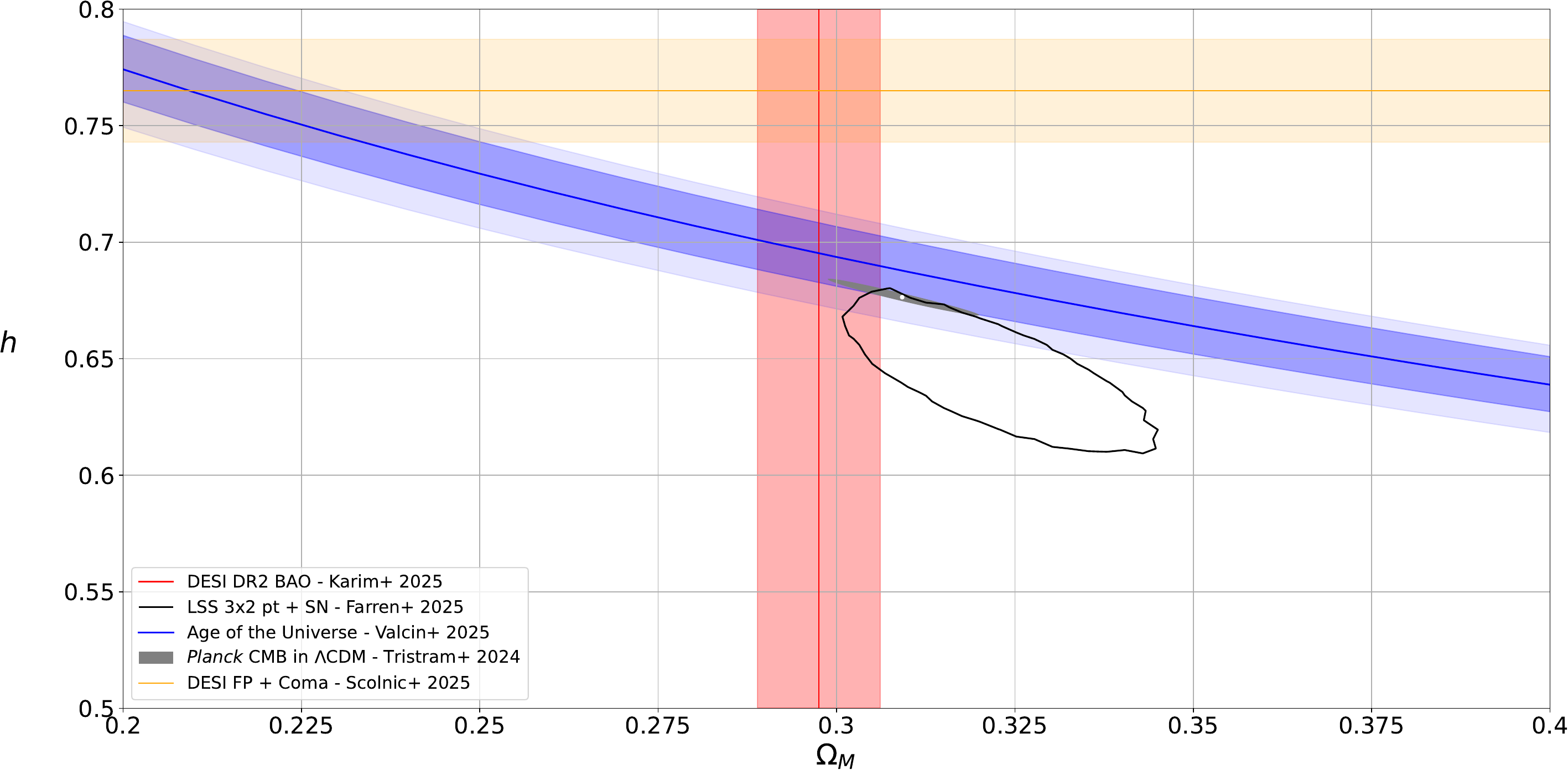}
    \caption{Similar to Figure~\ref{OmegaM_h}, but using different non-CMB constraints in each case. The shape of the expansion history is constrained using DESI~DR2 BAO measurements \citep[red;][]{DESI_2025}. The age of the Universe is estimated using Galactic GCs \citep[blue;][]{Valcin_2025}, with the same assumed formation time. The open black contour shows the constraint from uncalibrated SNe~Ia combined with three different 2-point statistics of large-scale structure, namely galaxy-galaxy, galaxy-lensing, and lensing-lensing \citep{Farren_2025}. The CMB is used as the source radiation, but the constraint largely derives from the clustering properties of matter at late times. This also contains information about the turnover scale in the matter power spectrum, and thus the epoch of matter-radiation equality (see the text). The yellow band shows $h$ estimated from the local redshift gradient using DESI~FP measurements \citep{Said_2025} calibrated using the distance to the Coma Cluster \citep{Scolnic_2025}. No measurement of its distance since 1990 exceeds the minimum of 110~Mpc required for consistency with the \emph{Planck} $\Lambda$CDM constraint.}
    \label{OmegaM_h_ACT_SN}
\end{figure*}

To check the robustness of this conclusion, we show an alternative version of each constraint in Figure~\ref{OmegaM_h_ACT_SN}. The overall picture remains very similar, with the $3\times2$pt analysis of LSS (Section~\ref{LSS}) preferring much the same region of parameter space as the overlap between the DESI~DR2 BAO constraint on $\Omega_{\mathrm{M}}$ (Section~\ref{UCS}) and the nearly orthogonal constraint from old Galactic GCs (Section~\ref{Age_constraint}). The region of consistency between all three probes also contains the very narrow region where \emph{Planck} CMB observations can be compatible with the $\Lambda$CDM paradigm. Once again, the only outlier is the local $cz'$, even though this time it is found using the DESI~FP relation of elliptical galaxies calibrated with the known distance to the Coma Cluster (Section~\ref{Local_redshift_gradient}).

The concordance of probes other than the local $cz'$ in both figures highlights the discrepant nature of the local $cz'$, which if equated with $H_0$ can only be reconciled with the observed age of the Universe at an unrealistically low $\Omega_{\mathrm{M}}$. We suggest instead that $cz'$ has been inflated by outflow from the local KBC supervoid \citep*{Keenan_2013}, which is evident across the whole electromagnetic spectrum, from radio to X-rays \citep*[see section~1.1 of][and references therein]{Haslbauer_2020}. Those authors estimated in their equation~5 that the apparent $46 \pm 6\%$ underdensity of the KBC void in redshift space implies the local $cz'$ should exceed the background $H_0$ by $11 \pm 2\%$, which is remarkably similar to the actual magnitude of the Hubble tension. The model developed by those authors solves the Hubble tension consistently with the observed void density profile. Without further altering the void model parameters, it also successfully predicted that the BAO data would deviate from expectations in the \emph{Planck} $\Lambda$CDM cosmology \citep{Banik_2025_BAO}. Those authors compiled a list of 49 BAO measurements over the last 20 years, excluding duplicates as far as possible to avoid double-counting (see their table~1). The \emph{Planck} cosmology gives a total $\chi^2$ of 93, but the void models reduce this to $55-57$, depending on the assumed density profile. This corresponds to reducing the tension from $3.8\sigma$ to just $1.1\sigma - 1.3\sigma$ (see their table~2). Since the parameters of both the \emph{Planck} cosmology and the void models were previously published without regard to BAO data and those parameters were used without adjustment to predict the BAO observables, this represents a direct comparison between the \emph{a priori} predictions of the two models. The data show a very strong preference for the local void model, which is hundreds of times more likely than the homogeneous \emph{Planck} cosmology even if we only consider BAO measurements from DESI~DR2 (see their section~3.2). Moreover, it can also fit the bulk flow curve out to $250/h$~Mpc \citep{Watkins_2023, Mazurenko_2024, Stiskalek_2025}. The model requires faster structure growth than expected in $\Lambda$CDM, as otherwise local structure cannot solve the Hubble tension through cosmic variance in $H_0$ \citep[expected to be only 0.9~km/s/Mpc;][]{Camarena_2018}. Independently of the proposed local void model, the fact that bulk flows on a scale of $250/h$~Mpc are $\approx 4\times$ the $\Lambda$CDM expectation for that scale \citep{Watkins_2023, Whitford_2023} implies that peculiar velocities are larger than expected by about this factor. This in turn would raise the expected cosmic variance in $H_0$ to perhaps 3~km/s/Mpc, which is sufficient to solve the Hubble tension as a $2\sigma$ local fluctuation, provided we live in an underdensity.

An interesting aspect of Figures~\ref{OmegaM_h} and \ref{OmegaM_h_ACT_SN} is that without the age constraint shown in blue, there is excellent agreement between the local $cz'$ and the constraints from LSS and UCS. This suggests an early time resolution to the Hubble tension that allows the CMB anisotropies to be fit at higher $h$. The problem is the absolute ages of old stars \citep{Cimatti_2023} and GCs \citep{Valcin_2025} -- and indeed the Galactic disc as a whole \citep{Xiang_2022, Xiang_2025}. While uncertainties in the absolute ages of stars and GCs might be underestimated \citep{Joyce_2023}, recent work suggests that spectroscopic ages should be fairly reliable \citep{Shariat_2025, Tomasetti_2025}. Moreover, it is possible to constrain cosmology using only relative stellar ages by treating stellar populations as cosmic chronometers \citep[CCs;][]{Moresco_2018, Moresco_2020}. The idea of the CC technique is that a `red and dead' galaxy formed its stars in a rapid burst at very early times, with the stellar population passively evolving ever since \citep{Jimenez_2002}. As stars of different masses have different lifetimes, we may observe stellar populations in such galaxies at different redshifts to determine the relative age between those redshifts. This is similar to observing the main sequence turnoff in a GC, but using relative rather than absolute ages. This can make the results less sensitive to evolutionary phases that might be hard to model, provided the model at least captures how the duration of such phases changes with stellar mass. The use of a much larger number of stars also reduces sensitivity to individual stars. The $dz/dt$ results obtained with CCs are almost entirely from $z > 0.3$, making them largely insensitive to the very low redshifts at which a local void might play a role \citep*{Mazurenko_2025}. This allows CCs to constrain the background cosmology in a somewhat different way to nearby stars. The slope of the age-redshift relation can only constrain some combination of $H_0$ and $\Omega_{\mathrm{M}}$, so \citet{Cogato_2024} combine CCs with uncalibrated SNe~Ia, BAOs, and $\gamma$-ray bursts (GRBs) treated as UCS, thereby obtaining $H_0 = 67.2^{+3.4}_{-3.2}$~km/s/Mpc. This is very consistent with the \emph{Planck} value, but the local $cz'$ is $\approx 2\sigma$ higher. More recently, \citet{Guo_2025} find that if the shape of the expansion history is constrained using UCS but the absolute timescale is constrained only from CCs, then $H_0 = 68.4^{+1.0}_{-0.8}$~km/s/Mpc, $4.3\sigma$ below the local $cz'$. In a similar vein, \citet{Cao_2023} report that $H_0 = 69.25 \pm 2.42$~km/s/Mpc by combining data from BAOs, CCs, SNe~Ia, GRBs, quasi-stellar object (QSO) angular sizes, and QSO reverberation mapping using the Mg~II and C~IV spectral lines. This result comes from averaging over the different models presented in their appendix~A, with statistical and systematic errors added in quadrature and HII galaxies excluded given they are presently not suitable for use in cosmology \citep{Cao_2024}. The good agreement between the constraints from CCs and from the absolute ages of ancient Galactic stars and GCs provides additional confidence that $H_0$ is below the local $cz'$, at least if we assume standard FLRW equations at $z \la 100$.

It is therefore difficult to solve the Hubble tension through modifications to standard physics at very early times, even if a good fit to the CMB can be retained \citep[a non-trivial task; see][]{Vagnozzi_2021}. Thus, the most promising way to retain the usual assumption that $cz' = H_0$ is to modify the expansion history at late times. This may have little impact on $t_{\mathrm{U}}$ and the angular diameter distance to the CMB. If the modification is carefully tuned, it could enhance the present expansion rate enough to solve the Hubble tension while not much affecting the other observables \citep{Tiwari_2024}. Such a modification may arise in scalar-tensor gravity theories \citep{Petronikolou_2023} or from a non-minimal coupling between curvature and gravity \citep{Barroso_2024}. Another possibility is that dark energy is not a pure cosmological constant \citep[e.g.,][]{Harko_2023, Yao_2023, Rezazadeh_2024}. The main problem with this is that a larger $H_0$ requires a higher energy density according to the Friedmann equations, which would mean that the dark energy density rises with time \citep{Dahmani_2023}. This leads to a phantom equation of state, which suffers from theoretical issues with vacuum stability \citep{Ludwick_2017} and violation of the null energy condition \citep{Lewis_2025}. These issues can be avoided in some proposals based on $f \left( R \right)$ or Horndeski gravity \citep{Montani_2024, Tiwari_2025}. Even so, it can be difficult to reconcile such models with the latest BAO data, which seems to imply that the dark energy density is currently decreasing with time, worsening the Hubble tension \citep{DESI_2025, Mirpoorian_2025}. This is probably related to the requirement that the angular diameter distance to recombination should not be modified, which in turn implies that the expansion rate can be faster than in the \emph{Planck} cosmology at some redshifts only if it is lower at other redshifts \citep[section~4 of][]{Banik_2025_BAO}.

Moreover, a purely background solution to the Hubble tension requires us to neglect the evidence that structure formation is more efficient than expected in $\Lambda$CDM, e.g. from the high redshift, mass, and collision velocity of the El Gordo interacting galaxy clusters \citep*{Asencio_2021, Asencio_2023}. A faster background expansion rate would create more Hubble drag on growing structures, making it more difficult to explain the fast observed bulk flows and form large and deep voids like the KBC supervoid \citep{Keenan_2013, Wong_2022}, whose observed redshift-space density contrast excludes $\Lambda$CDM at $6\sigma$ confidence \citep*{Haslbauer_2020}.

If these problems are overcome, a modification to the expansion history at late times would avoid the serious difficulties pointed out in this contribution with modifications at early times. These difficulties are unrelated to the issue of whether such modifications can achieve a good fit to the CMB anisotropies, which is far from certain \citep[see also][and references therein]{Vagnozzi_2023}. This issue is growing increasingly problematic because $\Lambda$CDM continues to fit the CMB anisotropies very well despite significant improvements to CMB observations, which by now pose significant problems for the most commonly considered early-time solutions to the Hubble tension \citep{Calabrese_2025, Camphuis_2025}.

\section{Conclusions}
\label{Conclusions}

We consider constraints on $H_0$ and $\Omega_{\mathrm{M}}$ from the shape of the cosmic expansion history as traced by SNe and BAO measurements without any absolute calibration of either \citep*[UCS;][]{Lin_2021_UCS}, the horizon size at matter-radiation equality as deduced from the turnover scale in the matter power spectrum \citep{Philcox_2022}, and the age of the Universe from old stars in the Galactic disc and halo \citep{Cimatti_2023}. These constraints assume a flat FLRW cosmology but are insensitive to the physics shortly prior to recombination, with the age and UCS constraints being almost completely immune to what happened in the first Myr. We find that the $1\sigma$ regions allowed by these constraints do not overlap, but a narrow region of parameter space is consistent with all these constraints just outside their $1\sigma$ confidence intervals (Figure~\ref{OmegaM_h}). Remarkably, the standard \emph{Planck} fit to the CMB anisotropies in $\Lambda$CDM falls precisely in this region \citep{Tristram_2024}. This strongly suggests that the anomalously high local $cz'$ (the Hubble tension) is not caused by a modification to $\Lambda$CDM at early times in cosmic history. This is in line with the excellent fit to the CMB anisotropies in $\Lambda$CDM despite ongoing improvements to CMB observations, which by now pose a problem for the most common early-time solutions to the Hubble tension \citep{Calabrese_2025, Camphuis_2025}.

We demonstrate the robustness of this conclusion by obtaining each non-CMB constraint from a different study (Figure~\ref{OmegaM_h_ACT_SN}). For this, we use uncalibrated DESI~DR2 BAO data \citep{DESI_2025}, a $3\times2$pt analysis of LSS \citep{Farren_2025}, and old Galactic GCs as a constraint on $t_{\mathrm{U}}$ \citep{Valcin_2025}. The overall picture remains very similar, with the only inconsistent measurement being the local $cz'$, which now comes from a different study compared to Figure~\ref{OmegaM_h}. The anomalously high local $cz'$ is evident in a great many recent studies on the issue, as discussed further in the extensive CosmoVerse white paper \citep{Valentino_2025} and references therein. There are several techniques that can be used in each rung of the distance ladder, but the various possible permutations give a numerically similar $cz'$ \citep{Scolnic_2023}. For instance, the role of Cepheids in the second rung of the distance ladder can be replaced by the tip of the red giant branch, while at the same time the role of SNe~Ia in the third rung can be replaced by surface brightness fluctuations, thereby achieving independence from the traditional Cepheid-SNe~Ia route to the local $cz'$ \citep{Jensen_2025}. It is also possible to use Mira variables in the second rung \citep{Bhardwaj_2025} or make do with Cepheids alone \citep{Stiskalek_2025_Cepheids}. Both studies report a value for $cz'$ that is in excellent agreement with other measurements but in considerable tension with the \emph{Planck} cosmology. Yet other techniques do not use any rungs, for instance Type~II SNe \citep{Vogl_2025} and megamasers \citep{Pesce_2020}. These also give a high local $cz'$ in line with the SH0ES result \citep{Breuval_2024}, which is based on using 4 anchor galaxies with geometric distances to calibrate the Leavitt Law \citep{Leavitt_1912}, itself used to estimate the distances to the host galaxies of 42 SNe~Ia and thereby fix the absolute magnitude scale of SNe~Ia in the Hubble flow ($z = 0.023-0.15$). Their result that $cz' = 73.17 \pm 0.86$~km/s/Mpc \citep{Breuval_2024} is now in $6.4\sigma$ tension with the expected value in $\Lambda$CDM calibrated to combined CMB observations from \emph{Planck}, ACT, and SPT \citep{Camphuis_2025}.

If this tension is not caused by previously unrecognised systematics that affect a wide variety of local distance indicators in a similar way, the required late-time modification to $\Lambda$CDM can either be at the background level through a slight tweak to the expansion history, or it could be due to a local inhomogeneity. Since cosmic variance in the local $cz'$ is only 0.9~km/s/Mpc in $\Lambda$CDM \citep{Camarena_2018}, solving the Hubble tension in this way would require enhanced structure formation on $\gtrsim 100$~Mpc scales, which might then permit the formation of a local supervoid that solves the Hubble tension. There is strong evidence for just such a void \citep*{Keenan_2013, Haslbauer_2020} and for bulk flows being about $4\times$ faster than expected in $\Lambda$CDM on scales relevant to the measurement of the local $cz'$ \citep{Watkins_2023, Whitford_2023}, which is to be expected in the local void scenario \citep{Mazurenko_2024, Stiskalek_2025}. This model also fits BAO observations over the last 20 years much better than $\Lambda$CDM \citep{Banik_2025_BAO}, avoiding the need for the dark energy density to evolve with time and undergo a theoretically problematic phantom crossing in the recent past. Moreover, the observed descending trend in the inferred value of $H_0$ with the redshift of the data used to measure it \citep{Jia_2023, Jia_2025a, Jia_2025b} agrees quite well with the \emph{a priori} predictions of the local void scenario \citep*{Haslbauer_2020, Mazurenko_2025}.

Our work adds to the growing evidence that modifications to $\Lambda$CDM at early times in cosmic history are unlikely to solve the Hubble tension, even if it is possible to preserve a good fit to the CMB anisotropies \citep{Vagnozzi_2023, Toda_2024}.

\section*{Acknowledgements} 
IB is supported by Royal Society University Research Fellowship 211046. NS is supported by the AKTION grant number MPC-2023-07755 and the Charles University Grants Agency (GAUK) grant number 94224. The authors thank Phillip Helbig and Harry Desmond for helpful comments. They are also very grateful to the anonymous referees for comments which significantly improved this manuscript.

\begin{appendix}

\section{The expansion history around matter-radiation equality}
\label{a_matter_radiation}

For the first several Gyr of the universe, the expansion history was almost entirely governed by the densities of matter and radiation. The Friedmann equation was thus of the form
\begin{eqnarray}
    \frac{\dot{a}}{a} ~=~ H_0 \sqrt{\Omega_{\mathrm{M}} \, a^{-3} + \Omega_{\mathrm{r}} \, a^{-4}} \, ,
    \label{FRW_matter_radiation}
\end{eqnarray}
where $a$ is the cosmic scale factor with value 1 today, an overdot denotes a time derivative, and $\Omega_{\mathrm{M}}$ ($\Omega_{\mathrm{r}}$) is the present fraction of the cosmic critical density in matter (radiation). We can solve this by separating the variables and using the substitution $a \equiv a_{\mathrm{eq}} \sinh^2 \theta$, where $a_{\mathrm{eq}} \equiv \Omega_r/\Omega_{\mathrm{M}}$ is the cosmic scale factor at matter-radiation equality. Applying the usual boundary condition that $a = 0$ when $t = 0$ gives the following result:
\begin{eqnarray}
    H_0 t \sqrt{\Omega_{\mathrm{M}}} ~=~ \frac{2 \, a_{\mathrm{eq}}^{3/2}}{3} \left[ 2 + \sqrt{\frac{a}{a_{\mathrm{eq}}} + 1} \left( \frac{a}{a_{\mathrm{eq}}} - 2 \right) \right].
    \label{t_a_relation}
\end{eqnarray}

Taking the ratio of $t$ at two different values of $a$ allows us to compute the ratio between the age of the universe at matter-radiation equality and recombination.
\begin{eqnarray}
    \frac{t_{\mathrm{rec}}}{t_{\mathrm{eq}}} ~=~ \frac{2 + \sqrt{\frac{a_{\mathrm{rec}}}{a_{\mathrm{eq}}} + 1} \left( \frac{a_{\mathrm{rec}}}{a_{\mathrm{eq}}} - 2 \right)}{2 - \sqrt{2}} \, ,
    \label{t_rec_eq_ratio}
\end{eqnarray}
where `eq' and `rec' subscripts refer to quantities at the time of matter-radiation equality and recombination, respectively. In $\Lambda$CDM, $a_{\mathrm{rec}} = 1/1090$ and $a_{\mathrm{eq}} \approx 1/3410$, so $a_{\mathrm{rec}}/a_{\mathrm{eq}} \approx 3.13$ \citep[table~2 of][]{Planck_2020}. This implies $t_{\mathrm{rec}}/t_{\mathrm{eq}} = 7.3$.

It is possible to invert Equation~\ref{t_a_relation} and obtain an expression for $a \left( t \right)$ by squaring both sides, which leads to a cubic expression for $a/a_{\mathrm{eq}}$. Substituting $u \equiv \left( a/a_{\mathrm{eq}} - 1 \right)/2$, we get an expression of the form
\begin{eqnarray}
    4 u^3 - 3 u ~=~ f \left( t \right),
\end{eqnarray}
for some function $f \left( t \right)$ that we do not repeat for clarity. This can be solved using a trigonometric triple angle formula by substituting $u = \cos \phi$, which yields
\begin{eqnarray}
    \cos \left( 3\phi \right) ~=~ f \left( t \right).
\end{eqnarray}
Combining these results, we get our final expression for the expansion history in the era when matter and radiation are the only appreciable components in the Friedmann equation:
\begin{eqnarray}
    \frac{a}{a_{\mathrm{eq}}} ~=~ 1 + 2 \cos \left[ \frac{1}{3} \acos \left( \frac{1}{2} \left[ \frac{3 H_0 t \sqrt{\Omega_{\mathrm{M}}}}{2 \, a_{\mathrm{eq}}^{3/2}} - 2 \right]^2 - 1 \right) \right].
    \label{a_t_relation}
\end{eqnarray}
Our derivation remains valid if the argument of the central $\acos$ function becomes $>1$, but then we need to use the generalised definition of $\cos$ and $\acos$ for complex numbers. This is equivalent to setting $\cos \to \cosh$ and $\acos \to \acosh$ in Equation~\ref{a_t_relation} and using only real quantities. If we instead use this equation as shown, the result of the $\acos$ operation smoothly transitions from the real axis to the imaginary axis of the complex plane, but the $\cos$ operation then returns us back to the real axis.

\section{Constraints excluding supernovae}
\label{Constraints_excluding_SNe}

\begin{figure*}
    \centering
    \includegraphics[width = \linewidth]{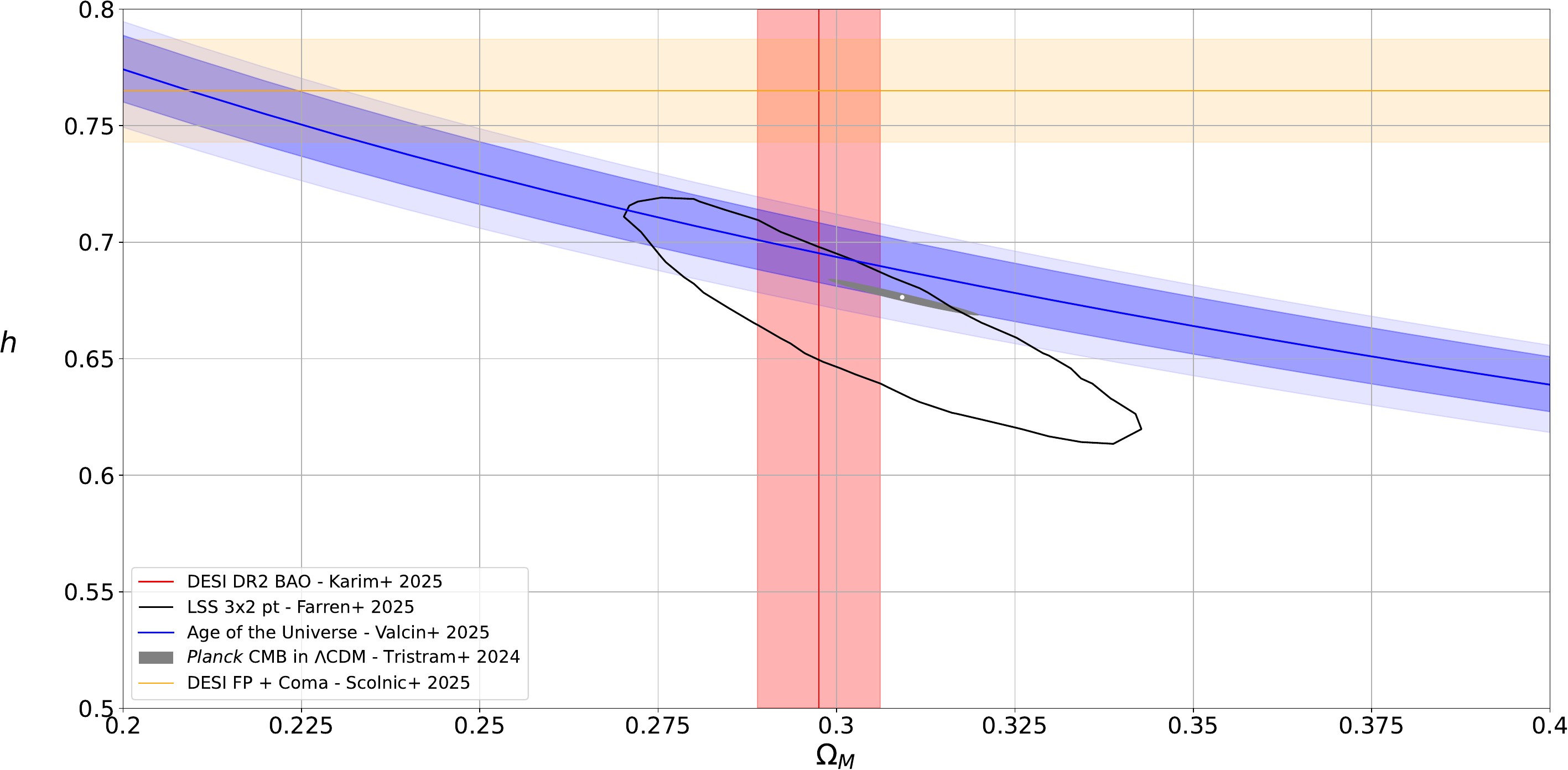}
    \caption{Similar to Figure~\ref{OmegaM_h_ACT_SN}, but with the LSS result \citep[open black contour;][]{Farren_2025} now shown excluding SNe~Ia.}
    \label{OmegaM_h_ACT}
\end{figure*}

SNe~Ia are often found at quite low redshifts, where any late-time or local solution to the Hubble tension would leave its mark. This makes it unwise to constrain cosmological parameters using SNe~Ia covering a wide redshift range. We therefore revisit Figure~\ref{OmegaM_h_ACT_SN} but exclude SNe~Ia from the \citet{Farren_2025} constraint, showing only their LSS $3\times2$pt result. This broadens the constraint, but does not qualitatively change the picture (Figure~\ref{OmegaM_h_ACT}). Comparing the constraints with and without uncalibrated SNe~Ia shows that these prefer a slightly high $\Omega_{\mathrm{M}}$. This is in line with analyses that only use uncalibrated SNe~Ia \citep{Brout_2022, DES_2024_SNe}.

\end{appendix}

\bibliographystyle{aasjournal}
\bibliography{Cosmology_no_CMB_bbl}
\end{document}